# Hydrodynamics of Semi-Submersible Vehicle Hulls with Variable Height-Width Ratio in Deep and Shallow Water

K.I. Matveev (Washington State University, USA)

**ABSTRACT**

Semi-submersible vehicles keep most of their hulls underwater while maintaining a small platform above the water surface. These craft can find use for both naval operations and civil transportation due to special properties, including the low above-water hull profile, reduced wave drag in some speed regimes, and potentially better seaworthiness. However, hydrodynamics of these marine craft is not well studied. In this work, computational modeling is undertaken to explore steady hydrodynamic characteristics of several semi-submersible hull variations in a range of speeds in deep-water and finite-depth conditions. The validation and verification study is conducted using experimental data obtained with a Suboff model in a near-surface regime. Parametric simulations are performed for this hull and two others generated by modifying the original Suboff geometry to produce narrow and wide hull shapes with similar volumes. The computational results indicate that the narrow hull excels in deep water, having lower drag and experiencing lower downward suction force and smaller longitudinal moment. However, in shallow-water operations, the narrow hull exhibits noticeably larger resistance than other hulls with the same displacement due to smaller gap between the hull and sea floor. Main hydrodynamic characteristics of the studied hulls and illustrations of wave patterns are presented in the paper. These findings can be useful for designers of semi-submersible vehicles.

**INTRODUCTION**

Semi-submersible vehicles (SSV) maintain most of their hulls underwater while keeping a small platform above the water (Figure 1). This makes SSV more difficult to detect in comparison with surface vessels. Small waterplane area (SWA) ships and submarines in surfaced states are the marine vehicle types closest to SSV (Dubrovsky, *et al.*, 2007; Burcher & Rydill, 1994). In contrast to submarines, SSV retain access to atmospheric air and are not intended for deep submersion. Due to reduced waterplane area, SSV are expected to have lower wave drag in some regimes and better seaworthiness. These properties of SSV are attractive for both clandestine military operations and commercial transportation at fast displacement speeds.

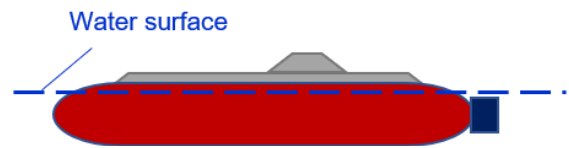

**Figure 1:** Schematic of a semi-submersible marine vehicle.

Technical publications on semi-submersible vehicles are rather limited. Construction and some test results of several concept SSV have been reported, ranging from large prototypes (Alleman, *et al.*, 2009) to medium-scale craft (Huo, *et al.*, 2021) to small-scale models (Spino & Matveev, 2023). Legacy knowledge available for submarines operating close to the free water surface (e.g., Burcher & Rydill, 1994) is useful but insufficient for comprehensive understanding of SSV hydrodynamics in a broad range of geometric variations and speed regimes. An exploratory study of design parameters for SSV has been recently presented by Sung et al. (2023).

Several papers specifically addressed hydrodynamics of mostly submerged hulls operating near the free water surface. Alvarez, *et al.*, (2009) reported results of their optimization study for hull shapes using a potential-flow method. Moonesun, *et al.*, (2016) employed a computational fluid dynamics (CFD) code to simulate a submarine-like model focusing on the effect of struts used in the towing test. Harwood, *et al.*, (2018) examined a bluff-bodied submersible and found that at shallow submersion, drag decreases at lower Froude numbers but increases at higher speeds relative to the surfaced position. Amiri, *et al.*, (2018) discussed importance of the interaction between the bow and aft shoulder waves produced by a submarine moving near the free surface. Matveev & Sung (2022) reported testing and modeling of a Suboff model (Liu & Huang, 1998) in the near-surface states; and hydrodynamics of a high-speed hard-chine semi-submersible craft was investigated numerically by Matveev (2022).

To provide more fundamental knowledge of SSV hydrodynamics, the present computational study addresses hydrodynamics of three basic semi-submersible hulls, one having the model-scale Suboff geometry (Liu & Huang, 1998) shown in Figure 2a,d, and two others that were generated by scaling the original hull form in the vertical and transverse directions. In one case (Figure 2b,e), a narrow hull was produced with the height increased by 20% and the beam decreased by 20%. In the other case (Figure 2c,f), a wide hull was generated by reducing the height by 20% and increasing the width by 20%.

These hulls are simulated in zero-trim and constant-sinkage conditions with the same static displacement, corresponding to the original hull's submergence-height ratio $h/H = 0.84$ (Figure 3). The motivation for using a narrow hull (Figure 2b,e) is to increase performance at relatively fast displacement speeds in deep water, when the wave drag may become very high. The wider hull (Figure 2c,f), although having inferior performance in deep water, can be more suitable for missions involving shallow-water operations, since detrimental finite-depth effects will be more pronounced for SSVs with large underwater hulls in comparison with conventional vessels.

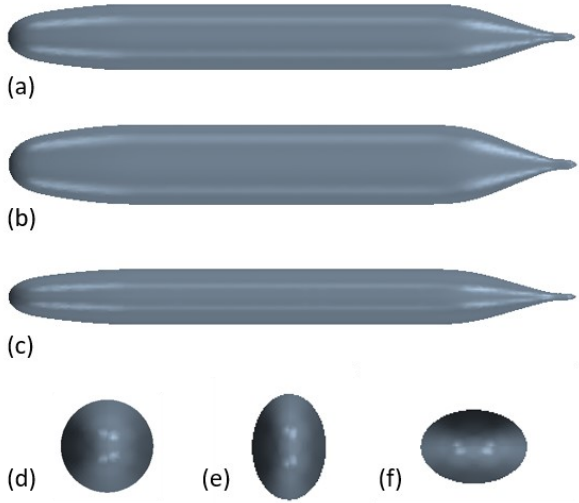

**Figure 2:** Three hull geometries investigated in this study: (a-c) side views, (d-f) front views. (a,d) Original Suboff hull, (b,e) narrow hull, (c,f) wide hull.

The computational approach employed in this work is discussed in the next section, followed by the verification and validation study. After that, parametric results for the lift, drag, and longitudinal moment and near-field wave patterns are reported for three semi-submersible hulls in a broad speed range in both deep and shallow water conditions.

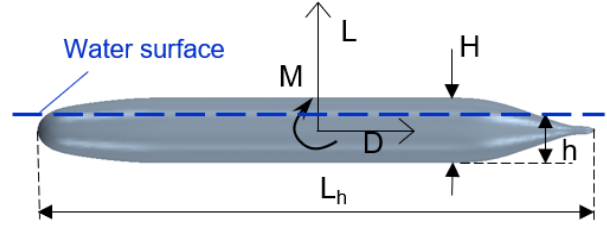

**Figure 3:** Schematic defining lift and drag forces, longitudinal moment, and submergence.

## COMPUTATIONAL MODELING APPROACH

Numerical simulations are conducted in this work using the computational fluid dynamics software Simcenter Star-CCM+. A finite-volume viscous solver with the 2$^{nd}$-order spatial and 1$^{st}$-order temporal discretization is employed. The results reported below correspond to steady-state conditions when time-averaged flow characteristics no longer evolve, so accurate modeling of unsteady flow features has not been attempted. The governing Reynolds-averaged Navier-Stokes equations (RANSE) used in this study involve the continuity and momentum equations (Ferziger & Peric, 1999),

$$\frac{\partial \rho}{\partial t} + \frac{\partial (\rho u_j)}{\partial x_j} = 0, \quad (1)$$

$$\frac{\partial (\rho u_i)}{\partial t} + \frac{\partial (\rho u_i u_j)}{\partial x_j} = -\frac{\partial p}{\partial x_i} + \rho f_i + \frac{\partial}{\partial x_j}\left[\mu\left(\frac{\partial u_i}{\partial x_j} + \frac{\partial u_j}{\partial x_i} - \frac{2}{3}\delta_{ij}\frac{\partial u_k}{\partial x_k}\right) - \rho \overline{u'_i u'_j}\right], \quad (2)$$

where $u_i$ is the Reynolds-averaged fluid velocity, $\rho$ is the fluid density, $p$ is the pressure, $f$ is the body (e.g., gravity) force, and $-\rho \overline{u'_i u'_j}$ is the Reynolds stress. To resolve gaseous and liquid phases, the volume-of-fluid (VOF) method is applied (Hirt & Nichols, 1981). The fluid density ρ and viscosity μ are calculated as $\rho = \rho_a \beta + \rho_w(1-\beta)$ and $\mu = \mu_a \beta + \mu_w(1-\beta)$, where $\beta$ is the volume fraction of air, and $a$ and $w$ stand for air and water, respectively. Water is treated as incompressible substance in the current simulations, whereas the ideal gas model is used for air. Gravity and surface tensions are also included in consideration.

The Reynolds turbulent stress is modeled with the Boussinesq hypothesis,

$$-\rho \overline{u'_i u'_j} = \mu_t \left( \frac{\partial u_i}{\partial x_j} + \frac{\partial u_j}{\partial x_i} - \frac{2}{3} \delta_{ij} \frac{\partial u_k}{\partial x_k} \right) - \frac{2}{3} \rho k \delta_{ij}, \quad (3)$$

where $\mu_t$ is the turbulent eddy viscosity and $k$ is the turbulent kinetic energy. The realizable $k - \varepsilon$ turbulence model is employed in this study (Rodi, 1991; Shih, et al., 1995), as it showed good performance for predicting surface flows in our previous investigations of hulls in various speed regimes (Matveev & Morabito, 2020; Wheeler, et al., 2021) and is frequently used for ship hydrodynamics CFD (De Luca, et al., 2016). The SST $k - \omega$ turbulence model (Menter, 1993) is tried as well in this work, producing similar results (within 1.5%). The governing transport equations for the turbulent kinetic energy $k$ and the turbulent dissipation rate $\varepsilon$ in the realizable $k - \varepsilon$ model are given as follows,

$$\frac{\partial(\rho k)}{\partial t} + \frac{\partial(\rho k u_j)}{\partial x_j} = \frac{\partial}{\partial x_j} \left[ \left( \mu + \frac{\mu_t}{\sigma_k} \right) \frac{\partial k}{\partial x_j} \right] + G_k - \rho \varepsilon, \quad (4)$$

$$\frac{\partial(\rho \varepsilon)}{\partial t} + \frac{\partial(\rho \varepsilon u_j)}{\partial x_j} = \frac{\partial}{\partial x_j} \left[ \left( \mu + \frac{\mu_t}{\sigma_\varepsilon} \right) \frac{\partial \varepsilon}{\partial x_j} \right] + \rho C_{\varepsilon 1} S \varepsilon - \rho C_{\varepsilon 2} \frac{\varepsilon^2}{k + \sqrt{\nu \varepsilon}}, \quad (5)$$

where $G_k$ is the production of turbulent kinetic energy by mean velocity gradients, $S = \sqrt{2 S_{ij} S_{ij}}$ is the scalar invariant of the strain rate tensor $S_{ij} = \frac{1}{2} \left( \frac{\partial u_i}{\partial x_j} + \frac{\partial u_j}{\partial x_i} \right)$, $\nu$ is the kinematic viscosity, $C_{\varepsilon 1}$ and $C_{\varepsilon 2}$ are the model coefficients, and $\sigma_k$ and $\sigma_\varepsilon$ are the turbulent Prandtl numbers for $k$ and $\varepsilon$, respectively. The turbulent eddy viscosity is determined by the following expression,

$$\mu_t = \rho C_\mu \frac{k^2}{\varepsilon}, \quad (6)$$

where $C_\mu$ depends on both the mean flow and turbulence properties (e.g., Mulvany, et al., 2004).

The computational domain has dimensions 3.7 x 2.1 x 1.8 of the hull length, which is equal to 1.894 m, as in the experiment used for validation (Matveev & Sung, 2022). The Suboff hull geometry is illustrated in Figure 2a,d. The boundary conditions in the numerical domain, shown in Figure 4, include velocity inlets on the front, top, and bottom boundaries, a pressure outlet at the downstream boundary, a slip wall on the port side, a no-slip wall on the hull surface, and a symmetry vertical plane passing through the hull centerplane. The numerical grid, consisting of hexahedral cells, is constructed inside the domain with refinements in the vicinity of the hull and free water surface (Figure 5). Prism layers are created near the hull to capture the boundary layer.

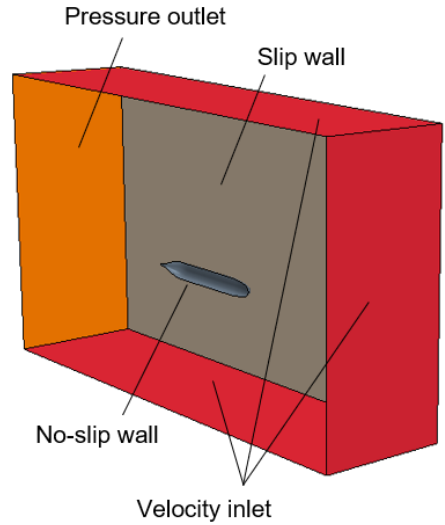

**Figure 4:** Numerical domain with boundary conditions.

The near-wall cell thicknesses were generally within 30-80 of Y+ values on the water-covered areas, which required the wall function approach. Due to the large variability of local Y+ values in multi-phase zones of the flow and the presence of dry sections of the hull, the blended approach based on the two-layer, all-Y+ method is utilized (Rodi, 1991).

**VERIFICATION AND VALIDATION STUDIES**

A mesh-dependency study was conducted with the Suboff hull to determine a suitable mesh density and to estimate numerical uncertainties. Three grids (coarse, medium, and fine) were generated, and simulations were carried out for the original hull at length Froude number 0.35, zero trim, and relative submergence $h/H = 0.84$ (Figure 3). The length Froude number is defined as follows,

$$Fr = \frac{U}{\sqrt{gL_h}}, \quad (7)$$

where $U$ is the hull speed, $g$ is the gravitational constant, and $L_h$ is the hull length.

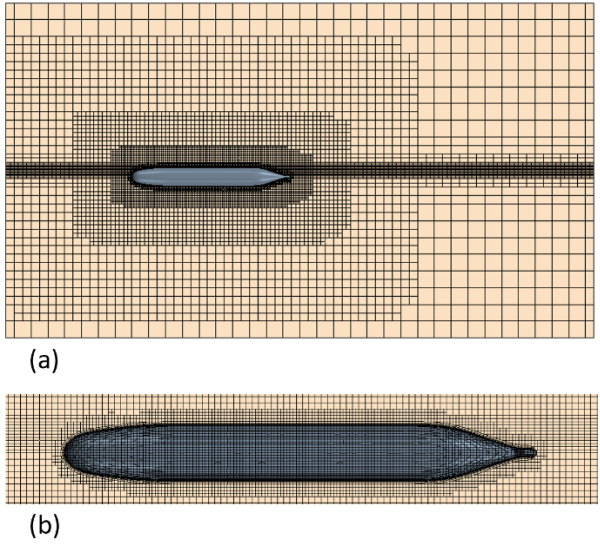

(a)

(b)

**Figure 5:** (a) Numerical grid in vertical (symmetry) plane of the domain, (b) mesh on the hull surface and in the hull vicinity, including free-surface zone.

As convergence metrics, drag $D$ and lift $L$ forces were used (Figure 3), which were also measured in the towing tests (Matveev & Sung, 2022; Sung, *et al.*, 2023). The lift comprises both hydrostatic and hydrodynamic forces. The results for the ratios of these forces to the static weight of the hull $W_0$ are shown in Table 1. The Richardson correction $\delta_{RE}$ (Ferziger & Peric, 1999) was computed and then multiplied by the safety factor $F$ to obtain the numerical uncertainty estimates $U_N$, following the standard procedure (Xing & Stern, 2010; ITTC, 2017),

$$\delta_{RE} = \frac{\Delta_{21}}{2^{p_{ob}} - 1}, \quad (8)$$

$$p_{ob} = \frac{\log(\Delta_{23}/\Delta_{12})}{\log(\beta)}, \quad (9)$$

$$U_N = F|\delta_{RE}|, \quad (10)$$

where $\Delta_{12}$ and $\Delta_{23}$ are the differences of solutions obtained on the fine and medium grids and the medium and coarse grids, respectively, $p_{ob}$ is the observed order of accuracy, and $\beta$ is the ratio of linear cell dimensions on different meshes. To evaluate the validation uncertainty $U_V$, the numerical ($U_N$) and experimental ($U_D$) uncertainties are combined as follows,

$$U_V = \sqrt{U_N^2 + U_D^2}. \quad (11)$$

**Table 1:** Numerical results for force ratios obtained on different grids and estimated uncertainties.

| Mesh | $D/W_0$ | $L/W_0$ |
|---|---|---|
| Coarse | 0.0253 | 0.8523 |
| Medium | 0.0286 | 0.8995 |
| Fine | 0.0290 | 0.9152 |
| Numerical uncertainty | 2.1% | 1.5% |
| Validation uncertainty | 2.4% | 1.8% |
| Difference between test and CFD results | 1.3% | 0.5% |

To validate the present numerical results, a comparison is made against experimental data obtained for the original hull at three Froude numbers (Matveev & Sung, 2022). Results for the drag-to-static-weight and lift-to-static-weight ratios are shown in Figure 6. The observed agreement is reasonably good. The differences between test and CFD results at $Fr = 0.35$ (used in the mesh-verification study) are found to be within the validation uncertainties (Table 1). Thus, the present numerical approach is deemed suitable for modeling of steady hydrodynamics of the considered semi-submersible hulls.

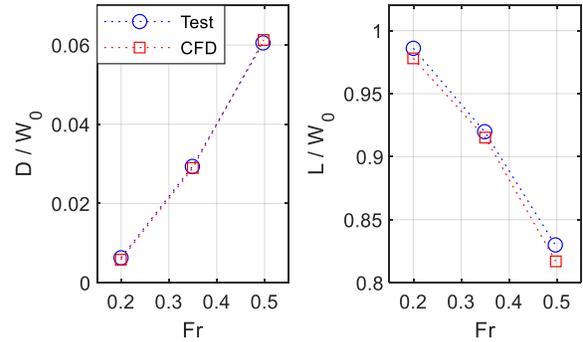

**Figure 6:** Comparison of experimental and numerical results for drag and lift forces normalized by static weight for Suboff hull (Matveev & Sung, 2022; Sung, *et al.*, 2023).

An additional validation case using similar numerical settings is conducted for another semi-submersible hull test reported in the literature (Moonesun, *et al.*, 2016; Matveev, 2022). The 1.3-m-long, 0.1-m-diameter model was towed at 80% submergence with fixed trim and heave. The hull geometry and an example of a calculated near-hull wave pattern near the hull are shown in Figure 7, where one can notice water build-up above the hull's bow section. A comparison between measured and numerically found resistance is given in Figure 8, showing good numerical prediction for this semi-submersible hull as well.

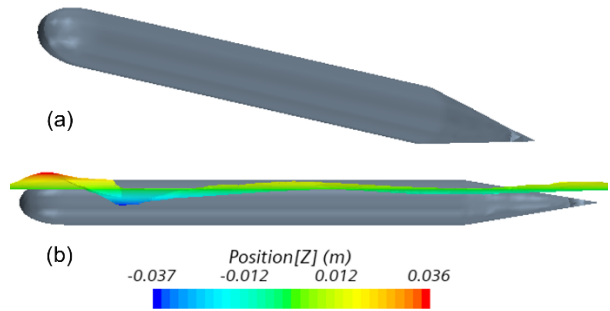

**Figure 7:** (a) Hull geometry (Moonesun, *et al.*, 2016); (b) near-hull wave pattern at speed 1 m/s.

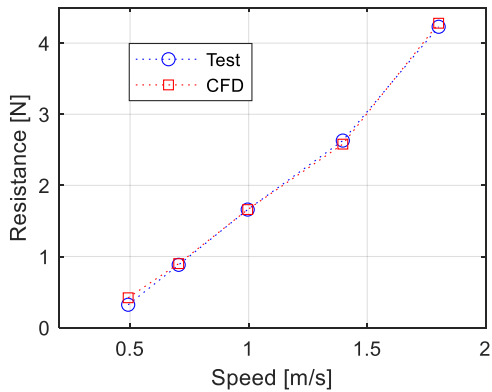

**Figure 8:** Comparison of test and simulation results for drag force (Moonesun, *et al.*, 2016; Matveev, 2022).

**RESULTS AND DISCUSSION**

In the first set of parametric simulations, three hulls (original, narrow, and wide, as shown in Figure 2) are simulated in deep water at length Froude numbers 0.2, 0.35 and 0.5, which for common surface vessels are usually associated with the displacement, fast displacement, and semi-displacement regimes, respectively. The hull attitudes are fixed at zero trim and relative submergence $h/H = 0.84$ that belongs to a typical range of semi-submersible vehicle sinkages.

The obtained drag and lift forces, as well as a moment about the transverse axis passing through the mid-ship point (Figure 3), are shown in Figure 9 in non-dimensional forms. The forces are normalized by the static displacement, whereas the moment is given relative to the static displacement multiplied by the hull length. (A bow-up moment is positive, and a bow-down moment would be negative.) Since the drag variation with speed is much greater than resistance differences between hulls at same speeds, drag forces of narrow and wide hulls normalized by drag of the original hull are also shown in Figure 9b to better display these differences. The water surface elevations for all nine cases (three hull forms at three speeds) are shown in Figures 10-12.

Resistances of all hulls drastically grow with Froude number increasing from 0.2 to 0.35 to 0.5 (Figure 9a); and much larger waves are generated at higher speeds (Figures 10-12). The narrow hull demonstrates drag lower than the original hull resistance by several percent (Figure 9b), as it more effectively pierces through the water. In contrast, the wide hull shows nearly symmetrical drag increase since it displaces more water in the vicinity of the free surface.

The overall lift of semi-submersible hulls significantly decreases with increasing Froude number (Figure 9c), as these hulls experience large downward suction forces due to accelerated water flow under the hulls, while also having water build-up on top of the hulls, which increases with speed (Fig. 10-12). The wide hull is most prone to lift reduction, as it has larger planform area that forces more water to go underneath and over the hull. The narrow hull is more favorable in this regard since it deflects incident water to the sides more effectively. This reduction of lift would need to be counteracted on free-to-heave hulls by fins or ballast reduction if one aims at maintaining the static sinkage.

The longitudinal moment shows a non-monotonic behavior with increasing speed (Figure 9d). Initially, up to $Fr = 0.35$, it decreases, indicating that the center of lift moves rearward. This is caused by more pronounced downward suction force under the front portion of the hull. However, at $Fr = 0.5$, the moment evolution reverses its trend and starts growing, as the hull enters the semi-displacement regime. Similar to fast boats in the pre-planing regime, hydrodynamic pressure builds up near the hull bow, while the suction force moves towards the stern. If the hulls were free to pitch, they would experience negative trim angles at moderate speeds and positive trims at higher speeds (Dubrovsky & Matveev, 2006). This sensitivity is a known problem for small-waterplane-area hulls, which can be counteracted either with fins (Lee & Curphey, 1977) or special bow forms (Kracht, 1999) or ballast re-distribution.

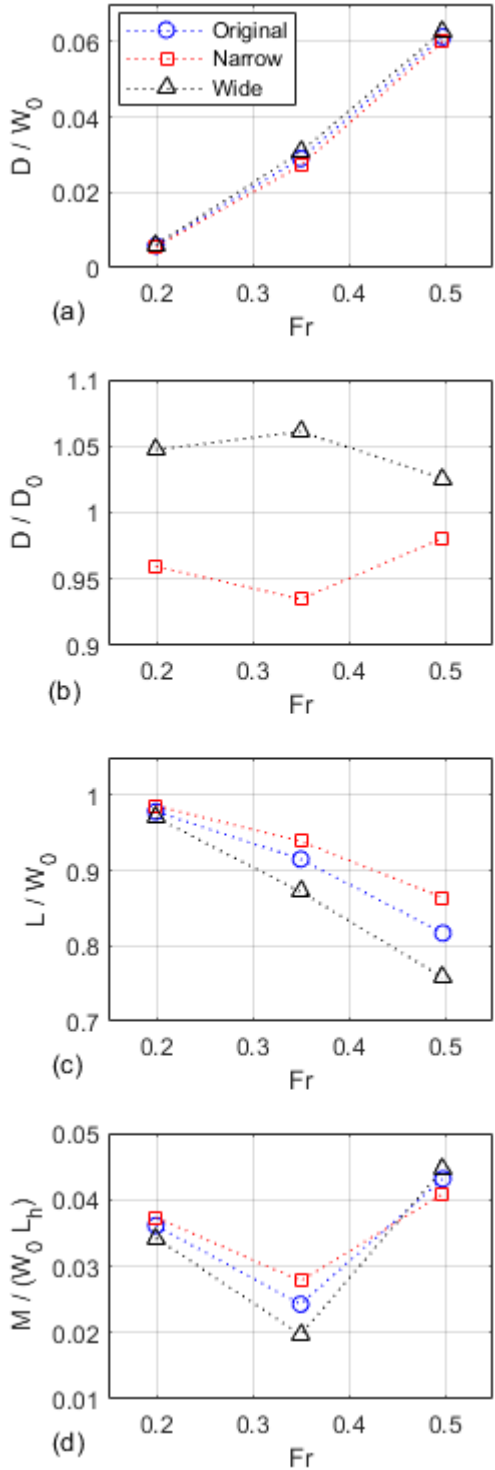

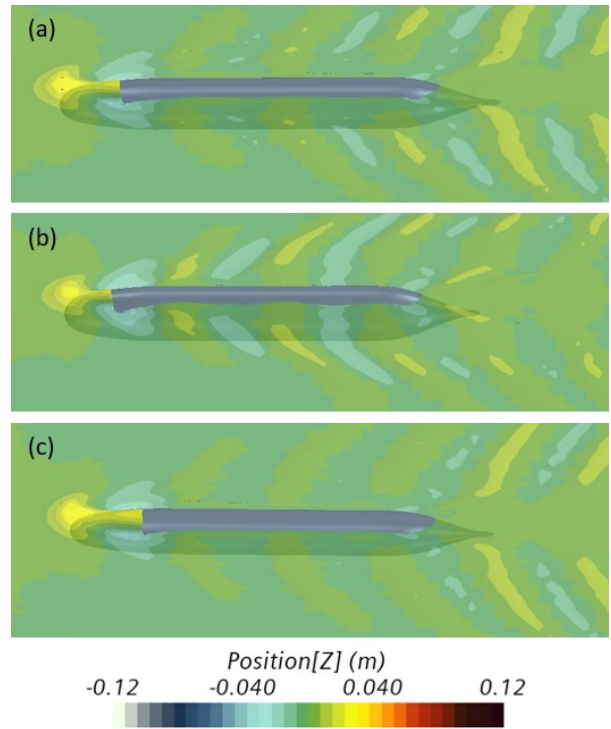

**Figure 10:** Water surface elevations at Fr = 0.20 in deep water: (a) original hull, (b) narrow hull, (c) wide hull.

**Figure 9:** Results for different hulls in deep water: (a) drag-to-static-weight ratio, (b) drag of narrow and wide hulls relative to drag of the original hull, (c) lift-to-static-weight ratio, (d) bow-up moment normalized by static weight and hull length.

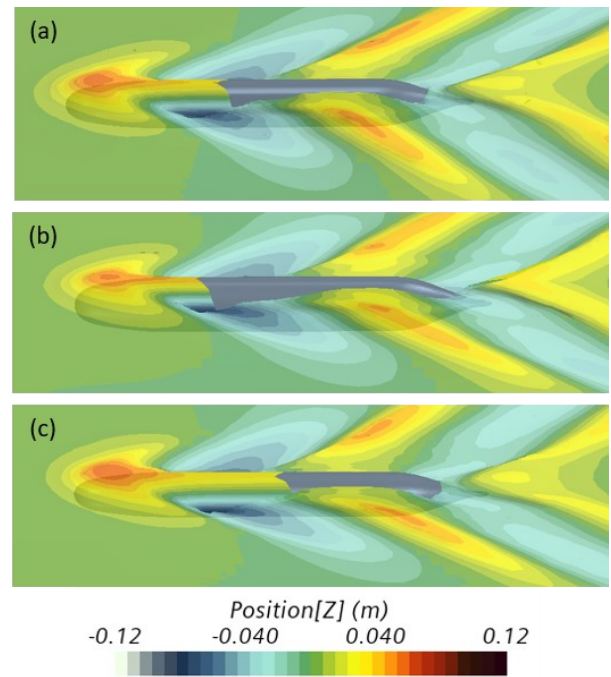

**Figure 11:** Water surface elevations at Fr = 0.35 in deep water: (a) original hull, (b) narrow hull, (c) wide hull.

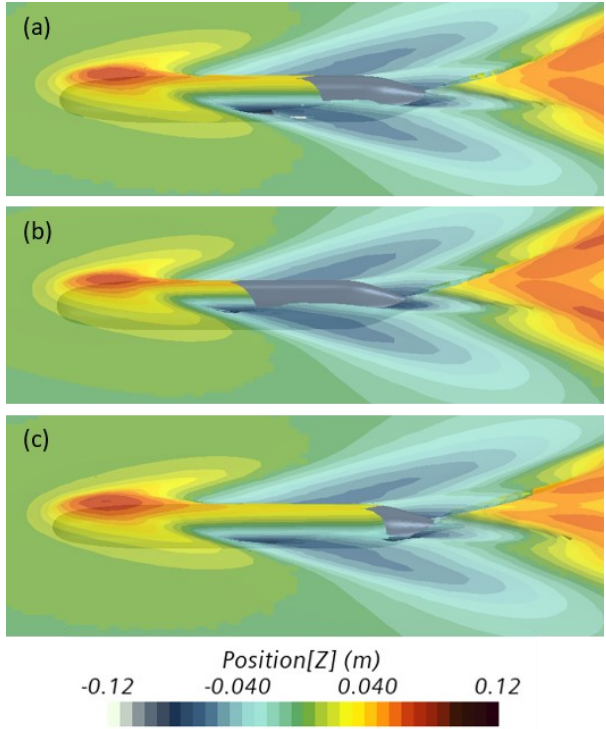

**Figure 12:** Water surface elevations at Fr = 0.50 in deep water: (a) original hull, (b) narrow hull, (c) wide hull.

In Figures 10-12, one can observe large increases of wavelengths and wave heights at higher Froude numbers, while the wave pattern variations between different hulls at the same speeds are relatively minor. The water build-up on top of bows and the extent of wetted zones along the hull top surfaces are more pronounced for the wide hull and less for the narrow hull.

Another series of simulations is conducted in very shallow water with the ratio of the water depth to the hull length being 0.168 (or the water depth to the original hull diameter ratio of 1.44). To appreciate the extreme shallowness of this condition, images of air fractions are shown in Figure 13 for both shallow- and deep-water cases of the narrow hull. One can notice much more pronounced water surface deformations in shallow water in comparison with the deep-water sailing at the same speed.

To describe shallow-water operations of surface vessels and especially relatively fast boats, the depth Froude number is often employed,

$$Fr_1 = \frac{U}{\sqrt{gH_w}}, \qquad (12)$$

where the water depth $H_w$ is utilized as a characteristic linear dimension instead of the hull length. Generally, in finite-depth situations with $Fr_1$ lower than one (subcritical regime), the hull resistance increases, while drag reaches a peak value when $Fr_1$ is close to one, i.e., in the so-called critical regime (Morabito, 2013). At larger $Fr_1$, resistance may drop below that in deep water at the same speed. However, in case of semi-submersible hulls, one may expect the critical regime to occur earlier, as relative draft of SSV is much larger than that of relatively fast surface boats. For the water depth used here, the depth-based Froude numbers $Fr_1$ were 0.49, 0.86 and 1.22, corresponding to the same length Froude numbers as in the deep water ($Fr$ = 0.2, 0.35, 0.5).

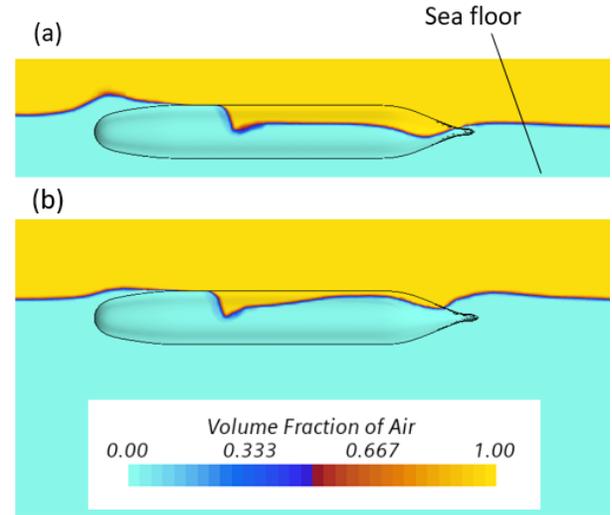

**Figure 13:** Volume fractions in the vertical symmetry plane and on the surface of narrow hull at Fr = 0.35: (a) shallow water, (b) deep water.

The shallow-water CFD results for the lift, drag, and moment normalized by the deep-water values are given in Figure 14, and the water surface elevations are illustrated in Figures 15-17. One can see that the finite-depth effects on the drag are relatively minor at the lower and higher Froude numbers, showing a few percent drag increase in the subcritical state and slight decrease in the supercritical regime. However, the resistances of the original and wide hull roughly double at $Fr = 0.35$ (or $Fr_1 = 0.86$) in comparison with the deep-water situation (Figure 14a).

Performance of the narrow hull with larger draft suffers even more with the drag rising more than three times. Thus, even if the wide hull has deep-water hydrodynamic properties somewhat inferior to other hull versions, this hull may be a better candidate for operations that include shallow water if resistance at sufficiently high (near-critical) speed and available water depth are important.

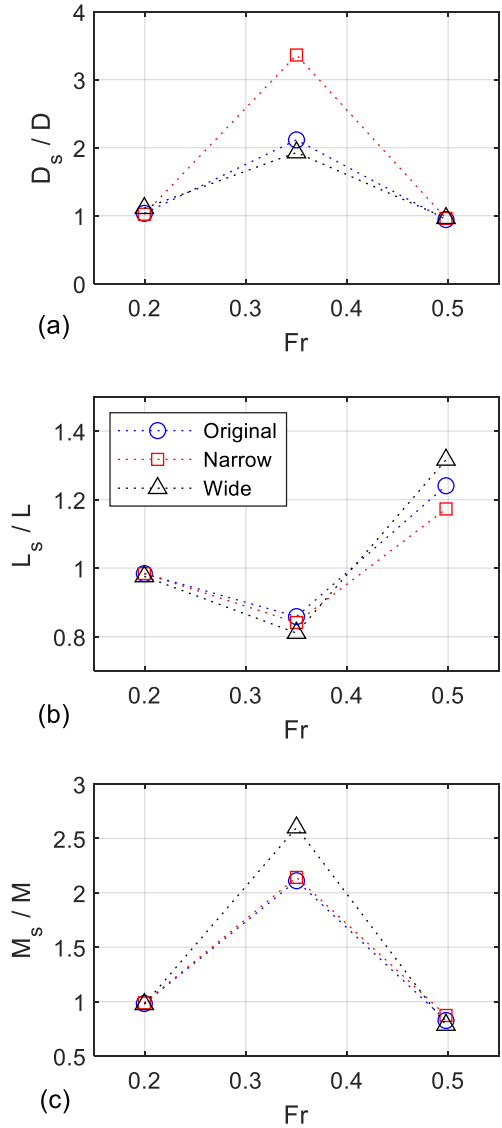

**Figure 14:** Shallow-water results normalized by values in deep water: (a) drag ratio, (b) lift ratio, (c) longitudinal moment ratio. Subscript s corresponds to shallow-water conditions.

However, the wide hull experiences slightly larger suction force increment in shallow water than the original and narrow hulls at $Fr = 0.35$, while generating more lift at $Fr = 0.5$ (Figure 14b). The longitudinal moment variation is also bigger for this hull at $Fr = 0.35$.

The water surface deformations in the finite-depth water at $Fr = 0.2$ (Figure 15) are almost the same as in the deep water (Figure 10), which is consistent with rather small differences between deep- and shallow-water hull forces and moments (Figure 14). However, at $Fr = 0.35$, the effect of finite depth on the water waves is noticeably larger for the original and wide hulls, and it is very big for the narrow hull (Figures 11 and 16). A stronger bow wave is formed near the original and wide hulls (Figure 16a,c), while depressed water surface with waves is visible around the aft portions of the hulls and downstream.

In case of the narrow hull with the largest draft, an even bigger soliton-like wave is observed near the bow and a more uniformly depressed region downstream of it (Figure 16b). In the supercritical regime at $Fr = 0.5$ (or $Fr_1 = 1.22$), large divergent waves originate at the bow and behind stern, and the overall water level is higher around hulls, but differences of the wave partners for different hulls become small (Figure 17). It should be noted that the current simulations with economical mesh and RANSE approach cannot fully capture flow unsteadiness that will accompany such large water surface disturbances.

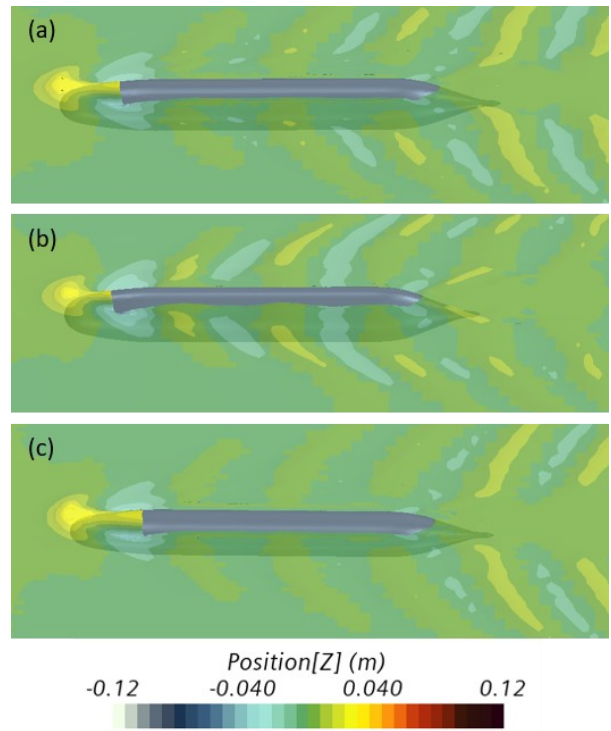

**Figure 15:** Water surface elevations at Fr = 0.20 in shallow water: (a) original, (b) narrow, (c) wide hulls.

Illustrations comparing velocity fields at the centerplane and the gage pressure distribution on the surface of the narrow hull moving in the deep and shallow water conditions are given in Figures 18 and 19. One can notice that in the shallow case there is more pronounced flow stagnation near the bow and faster flow under the rear part of the hull than in the deep-water case (Figure 18). This results in higher pressure on the front portion and reduced pressure in the rear bottom of the shallow hull (Figure 19a), while the pressure is more uniformly distributed along the hull moving in the deep water conditions (Figure 19b). This consequently leads to much larger drag and much higher bow-up moment of the narrow hull in the shallow-water situation (Figure 14a,c).

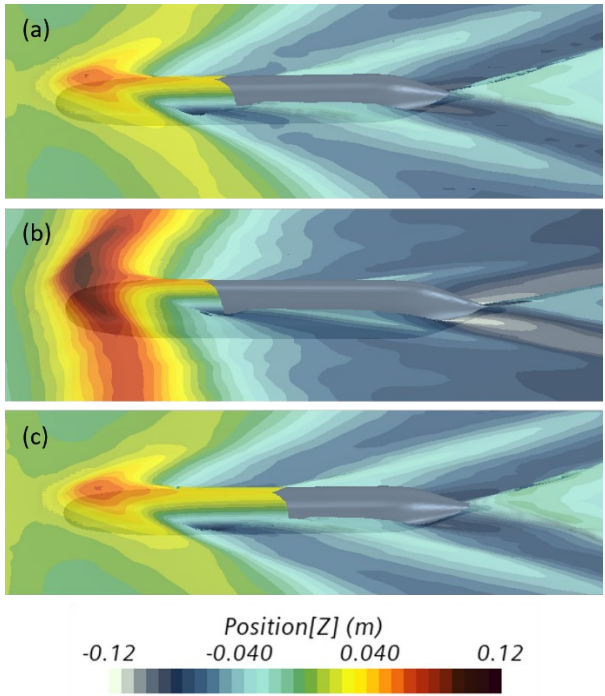

**Figure 16:** Water surface elevations at Fr = 0.35 in shallow water: (a) original, (b) narrow, (c) wide hulls.

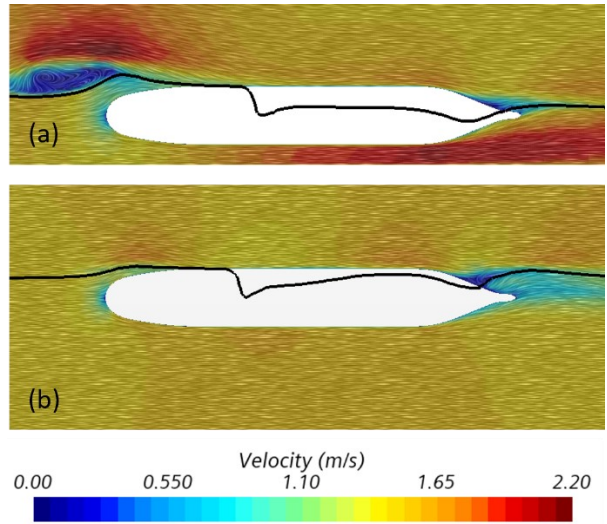

**Figure 18:** Velocity field in the vertical symmetry plane at Fr = 0.35: (a) shallow water, (b) deep water. Solid lines indicates the air-water interface at the centerplane and on the hull surface.

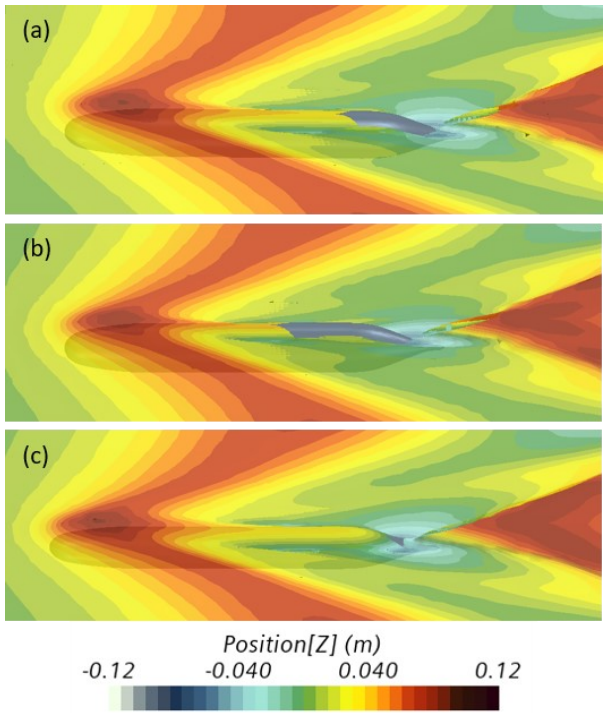

**Figure 17:** Water surface elevations at Fr = 0.50 in shallow water: (a) original, (b) narrow, (c) wide hulls.

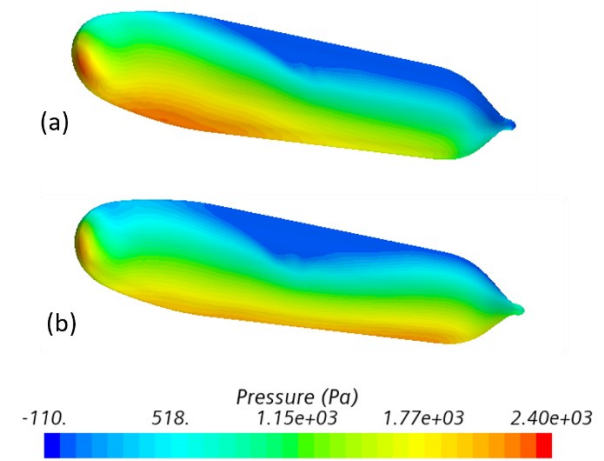

**Figure 19:** Pressure distribution on the narrow hull surface at Fr = 0.35: (a) shallow water, (b) deep water. Solid lines indicates the air-water interface at the centerplane and on the hull surface.

## CONCLUSIONS

In this work, CFD modeling has been applied to investigate steady hydrodynamics of semi-submersible hull variations at different speeds in deep and shallow water. Upon conducting the mesh-verification and validation analysis for a Suboff hull in the semi-submersible mode, narrower and wider versions having the same displacements were numerically modeled. In the deep water, the narrow hull showed several advantages, including lower drag, smaller downward suction force, and lower longitudinal moment, thus clearly being the most optimal choice among the studied hulls. However, in rather shallow water operations, this hull demonstrated noticeably larger drag increase than other hulls. The downward suction force increased for all hulls at $Fr = 0.35$ and decreased at $Fr = 0.5$. The shallow-water bow-up moment approximately doubles at $Fr = 0.35$ in comparison with the deep-water case, while it slightly decreases in the supercritical regime. Hence, if deep-water performance is of the most importance, then the narrow hull is certainly preferable, but in the limited-water-depth conditions, and especially at higher speeds, the original and wide hulls may be better candidates.

## DISCUSSION

Prof. Li P. Sung, Department of Naval Architecture & Ocean Engineering, United States Naval Academy, Annapolis, MD, USA.

Interesting analysis and results. One comment regarding the CFD simulations – it was implicit that the model was fixed in pitch and heave, which could be explicitly stated. As a result of the described dynamics effects on pitch and heave, it would be an interesting follow-on analysis to determine the steady state attitude. That would seem to be provide greater accuracy in resultant drag due to the realism of a free floating vessel.

In Figure 6, 7, 8 captions, it may be useful to remind the audience to the source of the test data and hull shape.

In figure 3, bow-up is shown as positive and on pages 7, 8, 9 in the text, you mention "bow-up moment" a couple times. However, in the Figure 14, it is described as "bow-down moment." This is a bit confusing or inconsistent.

Overall, great work and an exciting contribution to the hydrodynamics understand of low profile vessels, especially timely with the DoD interest in contested logistics solutions.

## AUTHOR'S REPLY

The author thanks Prof. Sung for a suggestion on the follow-up CFD study with a free-to-move hull and remarks on the bow-up moment description and references. The longitudinal moment has been described more carefully, and references to the test data sources have been added to the captions for each validation figure in the final version of the paper.

An additional set of CFD simulations has been carried out with the free-to-heave-and-pitch original hull at $Fr = 0.35$ and variable positions of the center of gravity (CG), with one of them corresponding to the equilibrium state of the fixed-attitude hull. The numerical simulation approach had to be changed to accommodate a moving hull. An overset mesh was formed around the hull, and fluid forces and body weight defined the body motion. The initial condition corresponds to the position used in the fixed-attitude study, whereas only steady-state results obtained after a transient period are reported here.

The dependencies of the steady-state hull submergence, trim angle, and drag on the CG location are presented in Figure A1. Two equilibrium hull positions with the center of gravity displaced forward and rearward from that of the zero-trim case are given in Figure A2.

One can observe a close correspondence between results of the fixed-attitude hull and the free-to-heave-and-pitch hull with the CG located at position predicted by the fixed hull's longitudinal moment and with the hull weight equal to the lift force of the fixed hull (Figure A1). A small difference in the results can be attributed to the difference in the numerical grids between these two situations, as a separate overset grid is generated around the hull, and the solution needs to be matched between two mesh regions. Thus, using the fixed hull with the attitude corresponding to the equilibrium state of the free floating hull is deemed acceptable in the present simulations.

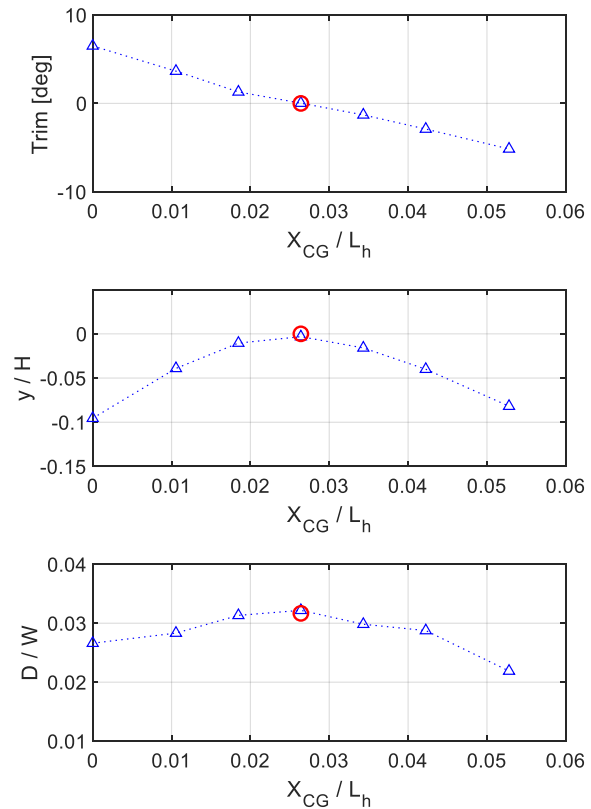

**Figure A1:** Dependence of the hull trim, relative submergence of the center of gravity (with respect to the fixed-hull position), and drag-to-weight ratio on the CG position for the original hull at Fr = 0.35. $X_{CG}$ is the location of the center of gravity forward from the hull mid-plane; $L_h$ is the hull length. Red circles indicate quantities obtained with the fixed-hull; blue triangles are the simulation results of the free-to-heave-and-pitch hull.

The parametric variation of the CG position leads to the expected behavior of the trim and submergence and an interesting effect on the drag force (Figure A1). As the CG moves forward from the zero-trim position, a negative trim is produced, while the rearward CG shift results in a positive trim. In both forward and rearward CG locations, the CG sinkage increases, as the trimmed hull buoyancy changes upon the rotation around the transverse line passing through CG. It is interesting to note that the drag force noticeably decreases at both positive and negative trims relative to the zero-trim case. This can be attributed to the reduction of the waterplane area and deeper submergence of the hull, which generally reduce the wave resistance. Despite achieving lower drag, utilization of a hull with a large steady-state trim would lead to a number of operational challenges, including positioning of the propulsion system.

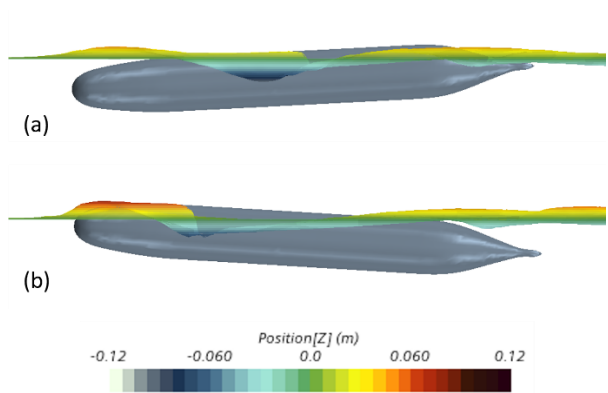

**Figure A2:** Steady-state positions of hulls and surrounding water surface with the center of gravity displaced (a) forward and (b) rearward from the zero-trim CG location by $\Delta X_{CG} / L_h = \pm 1.6\%$.